\documentclass[aps,pra,showpacs,english,10pt,a4,nofootinbib,notitlepage,twocolumn,superscriptaddress]{revtex4-1}

\usepackage[utf8]{inputenc}
\usepackage[T1]{fontenc}
\usepackage[english]{babel}  % english language

\usepackage{times}
\usepackage{dsfont}

\usepackage{amssymb,amsmath,amsfonts,amsthm}
\usepackage{mathtools}
\usepackage{verbatim}
\usepackage{enumerate}
\usepackage{graphicx}
\graphicspath{{figs/}}
\usepackage{hyperref}
\usepackage{bbm}
\usepackage{booktabs}

\usepackage[caption=false]{subfig}

\usepackage{color}
\definecolor{dred}{rgb}{.8,0.2,.2}
\definecolor{ddred}{rgb}{.8,0.5,.5}
\definecolor{dblue}{rgb}{.2,0.2,.8}
\definecolor{dgreen}{rgb}{.2,0.5,.2}

\newcommand{\bra}[1]{\mbox{$\langle #1|$}}
\newcommand{\ket}[1]{\ensuremath{|#1\rangle}}

\newcommand{\be}{\begin{equation}}
\newcommand{\ee}{\end{equation}}

\newcommand{\bea}{\begin{eqnarray}}
\newcommand{\eea}{\end{eqnarray}}

\begin{document}

\title{Quantum Simulation of Quantum Channels in Nuclear Magnetic Resonance}

\author{Tao Xin}
\thanks{These authors contributed equally to this work.}
\affiliation{State Key Laboratory of Low-Dimensional Quantum Physics and Department of Physics, Tsinghua University, Beijing 100084, China}
\affiliation{Tsinghua National Laboratory of Information Science and Technology,  Beijing 100084, China}

\author{Shi-Jie Wei}
\thanks{These authors contributed equally to this work.}
\affiliation{State Key Laboratory of Low-Dimensional Quantum Physics and Department of Physics, Tsinghua University, Beijing 100084, China}

\author{Julen S. Pedernales}
\affiliation{Department of Physical Chemistry, University of the Basque Country UPV/EHU, Apartado 644, 48080 Bilbao, Spain}
\affiliation{Institut f\"ur Theoretische Physik and IQST, Albert-Einstein-Allee 11, Universit\"at Ulm, D-89069 Ulm, Germany}

\author{Enrique Solano}
\affiliation{Department of Physical Chemistry, University of the Basque Country UPV/EHU, Apartado 644, 48080 Bilbao, Spain}
\affiliation{IKERBASQUE,  Basque  Foundation  for  Science,  Maria  Diaz  de  Haro  3,  48013  Bilbao,  Spain}

\author{Gui-Lu Long}
\email[Correspondence and requests for materials should be addressed to G.L.L.: ]{gllong@tsinghua.edu.cn}
\affiliation{State Key Laboratory of Low-Dimensional Quantum Physics and Department of Physics, Tsinghua University, Beijing 100084, China}
\affiliation{Tsinghua National Laboratory of Information Science and Technology,  Beijing 100084, China}
\affiliation{The Innovative Center of Quantum Matter, Beijing 100084, China}

\begin{abstract}
We propose and experimentally demonstrate an efficient framework for the quantum simulation of quantum channels in Nuclear Magnetic Resonance (NMR). Our approach relies on the suitable decomposition of non-unitary operators in a linear combination of $d$ unitary ones, which can be then experimentally implemented with the assistance of a number of ancillary qubits that grows logarithmically in $d$. As a proof-of-principle demonstration, we realize the quantum simulation of three quantum channels for a single-qubit: phase damping (PD), amplitude damping (AD), and depolarizing (DEP) channels. For these paradigmatic cases, we measure key features, such as the fidelity of the initial state and the associated von Neumann entropy for a qubit evolving through these channels. Our experiments are carried out using nuclear spins in a liquid sample and NMR control techniques.   
\end{abstract}

\maketitle

\section{Introduction}
In the last decades, significant progress has been achieved in the isolation and coherent control of quantum systems, allowing for the observation of their unitary dynamics~\cite{Lloyd,N,QS1,fenggrprl,QS2,fenggrsrep,QS3}. Such a degree of controllability has resulted in the implementation of quantum machines composed of a growing number of qubits, which have been used for key tests of quantum simulations and quantum computers.  As envisioned by Richard Feynman~\cite{Fey82}, large-scale quantum simulators would open the door to the analysis of new quantum physical phenomena and to the study of various models that are nowadays intractable with classical computers. In opposition to quantum simulators of closed quantum systems,  the simulation of open quantum systems, which has also been the subject of some research both from a theoretical~\cite{SL2001, ED2008,PRB2017,PRA2017,ARXIV2017, wang2013solovay,SC,sweke2014simulation, DiCandia15} and experimental~\cite{helu,N2011,NP2013} point of view, has been comparatively less explored. In this sense, both from a theoretical and experimental perspective, simulating open quantum systems pose relevant challenges. For example, understanding how quantum systems interact with their environment could potentially shed light on the physics of photosynthetic processes or transport phenomena in general~\cite{Huelga13, Mostame12}, which in turn could help design more efficient light-harvesting devices~\cite{Scully11, Dorfman11,Creatore13}. It could also help understand dissipation and thermalisation processes, or the nature of phase transitions. In the same manner, topics related to the foundations of quantum physics, as the measurement process or the quantum-to-classical transition~\cite{Zurek03}, would greatly benefit from a deeper physical understanding of open quantum systems.

In this work, we consider the simulation of a general CPTP channel dynamics and provide an efficient quantum algorithm for the implementation of non-unitary quantum dynamics associated to paradigmatic quantum channels. Our approach works by decomposing the non-unitary operators into a linear combination of unitary ones. This can be physically implemented via the assistance of a number of ancillary qubits that scales logarithmically with respect to the number of the involved unitary operators. We experimentally demonstrate our proposed quantum simulation method via the implementation of a set of decoherence quantum channels on a nuclear spin-qubit with NMR control techniques. More specifically, we implement the phase damping (PD), the amplitude damping (AD), and the depolarising (DEP) channels.

\section{Theoretical Results}

 An open quantum system can be defined as a subsystem of a larger system that includes the open system and its environment and follows a unitary dynamics, as described by $\rho_{\text{se}} =U(\rho \otimes \rho_{\text{env}})U^{\dagger}$. Here,  $\rho$ and $\rho_{\text{env}}$ are the initial states of the system and the environment, respectively, and are considered to be initially uncorrelated. The evolution of the principal system can be retrieved as $\rho_{\text{s}}=\text{tr}_{\text{env}}(U(\rho \otimes \rho_{\text{env}})U^{\dagger})$, where tr$_{\text{env}}$ is the partial trace over the environment degrees of freedom~\cite{N}. Alternatively, the evolution of the system  can also be described by a completely positive and trace-preserving map~\cite{ruskai2002analysis}: $\large \varepsilon (\rho)=\sum_{k}E_{k}\rho E_{k}^{\dagger}$, where $E_{k}$ are Kraus operators satisfying $\sum_{k}E_{k}^{\dagger}E_{k} =I $. Non-unitary processes of open quantum systems can also be described by master equations. While the Kraus formalism provides the description of the dynamics for a discrete time step, a master equation can provide a continuous time evolution of the density matrix that describes the open quantum system. 

Our method builds upon the framework of the so-called duality quantum computing (DQC) \cite{r1}. Such a framework allows for the arbitrary sum of $d$ unitary operators acting on an $n$-qubit system by the addition of log$_2(d)$ two-level ancillary systems.  Considering that Kraus operators  \{$E_k$\} can also be decomposed into a linear sum of  $d$ unitary operators, DQC appears to be of direct applicability to the simulation of open quantum system. A schematics of our proposal follows these steps:\\
\textbf{($a$)} A $d$-dimensional ancillary system is added to our working system (for example by the addition of $n=\log_2 d$ qubits) and the setup is initialised in the state $|\Psi\rangle|0\rangle$, where $|\Psi\rangle$ and $|0\rangle$ are the input states of the working system and the $d$-dimensional ancilla, respectively.  One additional operation $V$ is then performed on the auxiliary qudit $|0\rangle$,  transforming the system to: $|\Psi\rangle|0\rangle \rightarrow \sum_{i=0}^{d-1}V_{i0}|\Psi\rangle|i\rangle$, where $V_{i0}$ are the first column elements of the unitary matrix $V$ and are determined by the target map \{$E_k$\}.\\
\textbf{($b$)} The controlled operation $U_c= U_0\otimes \ket{0} \bra{0}+U_1 \otimes \ket{1}\bra{1}+....+U_{d-1}\otimes  \ket{d-1}\bra{d-1}$  is implemented afterwards. Here, $U_0$, $U_1$, ... , $U_{d-1}$ are the unitary basis corresponding to the decomposition of the elements \{$E_k$\}. This will result in the system evolving to the state $\sum_{i=0}^{d-1}V_{i0} U_i|\Psi\rangle |i\rangle$.\\
\textbf{($c$)}  Operation $W$ is performed on the auxiliary system, resulting in $\sum_i V_{i0} U_i|\Psi\rangle W|i\rangle=\sum_i\sum_k W_{ki} V_{i0} U_i|\Psi\rangle |k\rangle$, where  $W_{ki} V_{i0}$ are complex coefficients, and the sum $\sum_{i=0}^{d-1} W_{ki} V_{i0}=(WV)_{k0}$ corresponds to the $(k,0)$ element of the unitary matrix $WV$, and therefore satisfies $|\sum_{i=0}^{d-1} W_{ki} V_{i0}|\le 1$.  Thus, given a non-unitary transformation described by \{$E_k$\}, its corresponding evolution can be efficiently implemented if the unitary operations $V$, $W$, and $U_c$, satisfying $ E_{k}= \sum_i W_{ki} V_{i0} U_i$, are found. Notice that the first column of $V$ is defined by the specific decomposition of the Kraus operators into unitary operators that is chosen, while the rest of the matrix can be arbitrarily completed, with the only requirement of it being unitary. On the other hand, matrix $W$ is uniquely determined by $V$.\\
\textbf{($d$)} Finally,  measuring  the corresponding final state of the working system, with the ancillary system in state $\ket{k}\bra{k}$, will result in $E_k\ket{\Psi}\bra{\Psi}E^{\dagger}_k$. Therefore, if we trace out from the final state of the complete system, the degrees of freedom associated to the ancillary qubits, that is, if we sum over each state $\ket{k}\bra{k}$, with $\{ | k \rangle \}$ a complete basis of the ancillary system,  the result $\large \varepsilon (\rho)=\sum_{k}E_{k}\rho E_{k}^{\dagger}$, with $\rho=\ket{\Psi}\bra{\Psi}$, will correspond to the simulation of the map \{$E_k$\}.
\begin{figure}[htb]
\begin{center}
\includegraphics[width= 0.8\columnwidth]{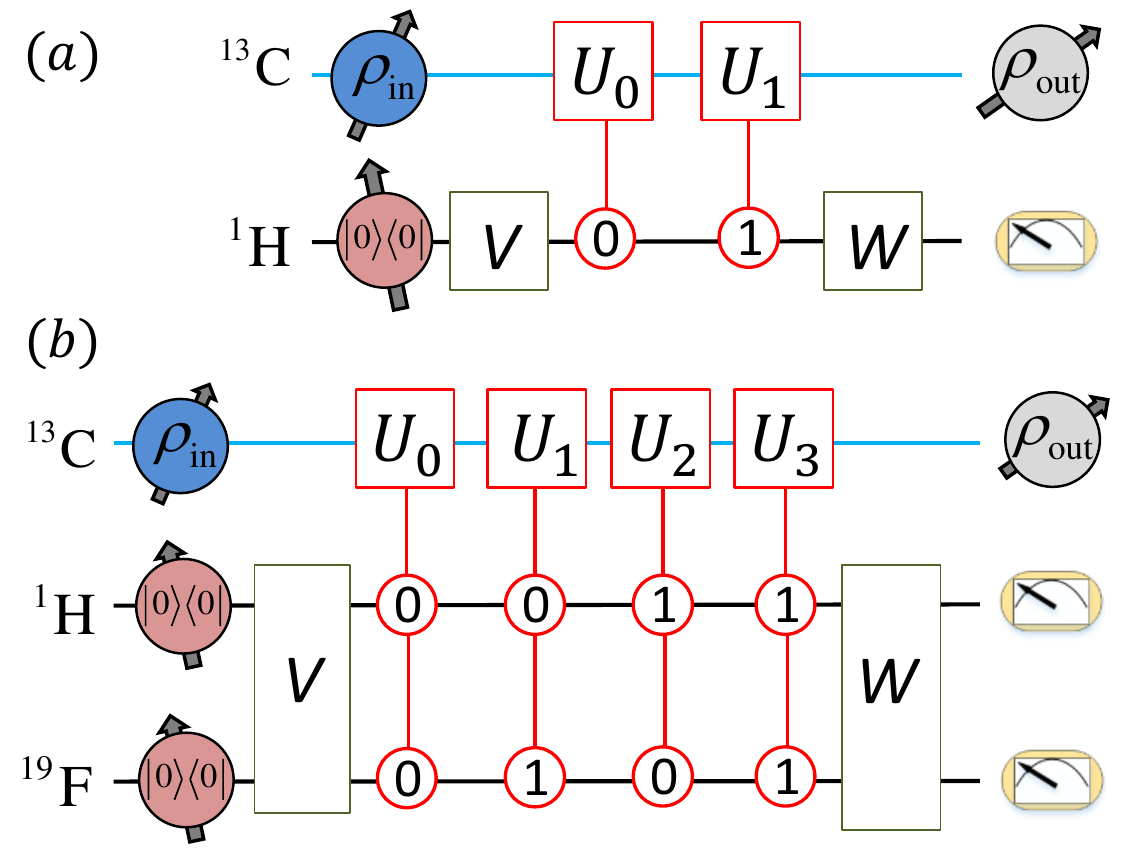}
\end{center}
\setlength{\abovecaptionskip}{-0.00cm}
\caption{\footnotesize{\textbf{Quantum circuit for the realisation of the PD (AD) channel (a), and the DEP channel (b).} The black lines are the ancilla (held by the nuclear spins of $^{13}$H and $^{1}$F), and the blue line is the system qubit (held by the nuclear spin of $^{19}$C). The red blocks represent the controlled operations. Operation $U_i$ is applied on the system qubit if the ancilla qubits are in the state $\ket{i}$, with $i=0, 1, ..., d-1$. }}\label{circuit1}
\end{figure}

\section{The three paradigmatic channels and experiments}
{\it{PD channel .--}} We will start the illustration of our method analysing the effect of a PD channel acting on a single-qubit~\cite{walls1985effect}. The effect of the PD channel is to remove the coherences of the qubit stored in the non diagonal elements of its density matrix~$\rho_{\text{in}}$.
In the Kraus representation, this corresponds to $E_0=[1~0;0~\sqrt{1-\lambda} ]$ and $E_1=[0~0;0~\sqrt{\lambda}]$, where the parameter $\lambda\in[0,1]$ represents the strength of the PD channel. In Fig. \ref{circuit1}(a) we give the quantum circuit that would realize such a noise channel according to the method introduced in this paper, which needs the addition of a single ancillary qubit.  For this case, Kraus operators $E_0$ and $E_1$ can be decomposed into  a linear combination of the unitary operators $\mathcal{I}$ and $\sigma_z$, where $\mathcal{I}$ is a $2\times2$ identity matrix and $\sigma_{x,y,z}$ are Pauli matrices. The decomposition is given by $E_0=\frac{1+\sqrt{1-\lambda}}{2}\mathcal{I}+\frac{1-\sqrt{1-\lambda}}{2}\sigma_z$ and $E_1=\frac{\sqrt{\lambda}}{2}\mathcal{I}-\frac{\sqrt{\lambda}}{2}\sigma_z$.

It can be easily checked that the unitary operators $V$, $W$, $U_0$ and $U_1$ that fulfil conditions $E_k=\sum_{i=0}^{1}W_{ki}V_{i0}U_i$ $(k=0,1)$ for a PD channel are given by
\begin{equation}
\begin{array}{l}
U_0=\mathcal{I}, U_1=\sigma_z,V=W=\left( {\begin{array}{*{20}{c}}
\sqrt{\frac{1+\sqrt{1-\lambda}}{2}}&\sqrt{\frac{1-\sqrt{1-\lambda}}{2}}\\
\sqrt{\frac{1-\sqrt{1-\lambda}}{2}}&{-\sqrt{\frac{1+\sqrt{1-\lambda}}{2}}}
\end{array}} \right).
\end{array}
\label{vw1main}
\end{equation}
As illustrated in Fig. \ref{circuit1}(a), the composite system consisting of an ancillary qubit and a working qubit is initialised in state $\rho^{\text{CH}}_{\text{in}}=\rho_{\text{in}}\otimes\ket{0}\bra{0}$,  with the input state of the working qubit $\rho_{\text{in}}=\ket{\phi}\bra{\phi}$. In order to extract the evolution corresponding to the PD channel acting on the working qubit, we need to trace out the ancillary degrees of freedom from the final state~$\rho_{\text{out}}^{\text{CH}}$. After doing so, the final state of the working qubit should correspond to $\rho_{\text{out}}=\large \varepsilon^{\text{PD}}(\rho_{\text{in}})=E_0\rho_{\text{in}}E_0^\dag+E_1\rho_{\text{in}}E_1^\dag$. The subspace where the ancillary qubit is in the state $\ket{0}$ will be associated with the evolution of the working system that corresponds to $E_0\rho_{\text{in}}E_0^\dag$, while the subspace of the ancilla state $\ket{1}$ will be associated to $E_1\rho_{\text{in}}E_1^\dag$.

\begin{figure}[htb!]
\begin{center}
\includegraphics[width= 0.98\columnwidth]{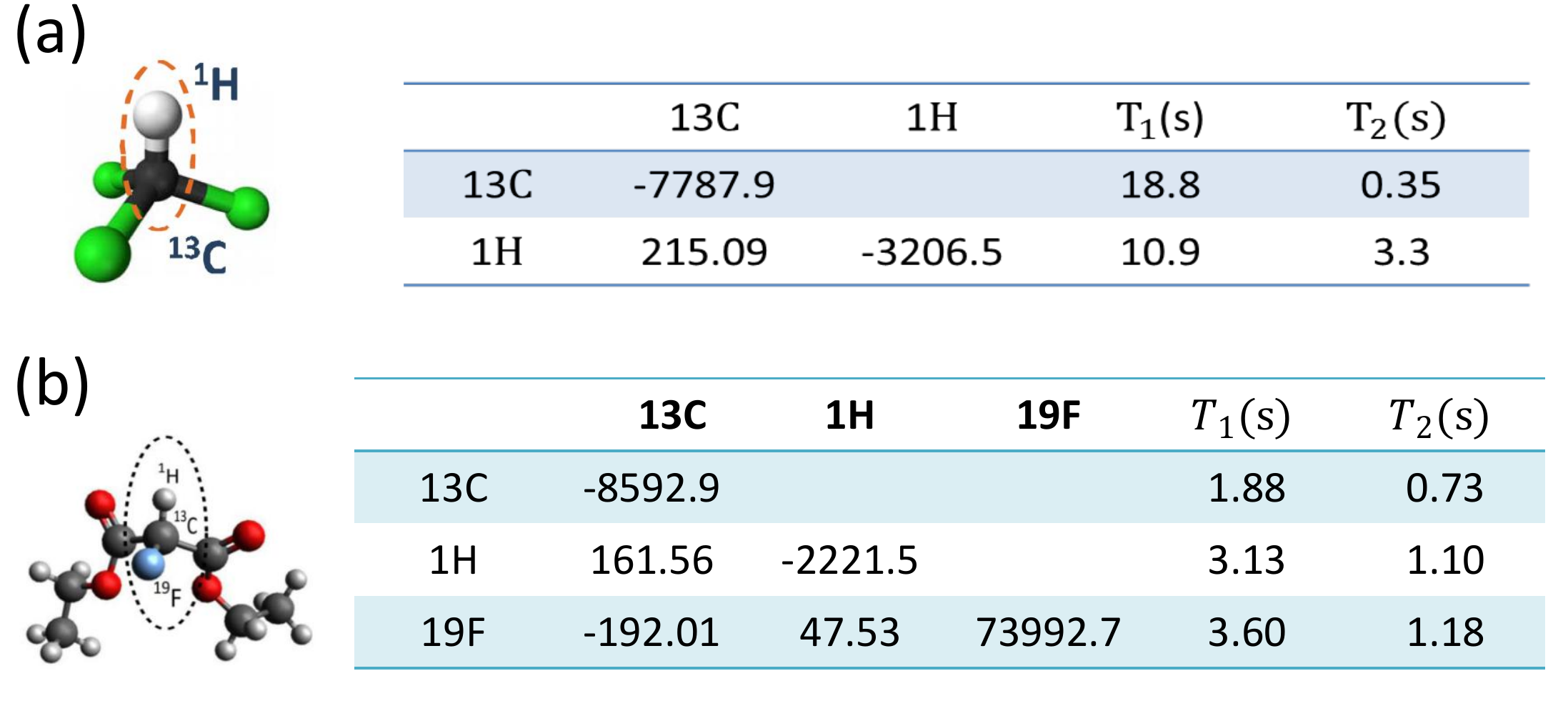}
\end{center}
\setlength{\abovecaptionskip}{-0.00cm}
\caption{\footnotesize{\textbf{Molecular structure and relevant parameters of experimental samples. (a) $^{13}$C-labeled Chloroform. (b) The diethyl fluoromalonate.}  The values of the chemical shifts (Hz) and $J$-coupling constant (Hz) between the column and row nuclei of the molecule are represented by the diagonal and off-diagonal elements of the table, respectively. The tables also provide the longitudinal time $T_1$ and transversal relaxation $T_2$, which can be measured using the techniques such as the standard inversion recovery.}}\label{molecule}
\end{figure}
\begin{figure*}[htb]
\begin{center}
\includegraphics[width= 1.7\columnwidth]{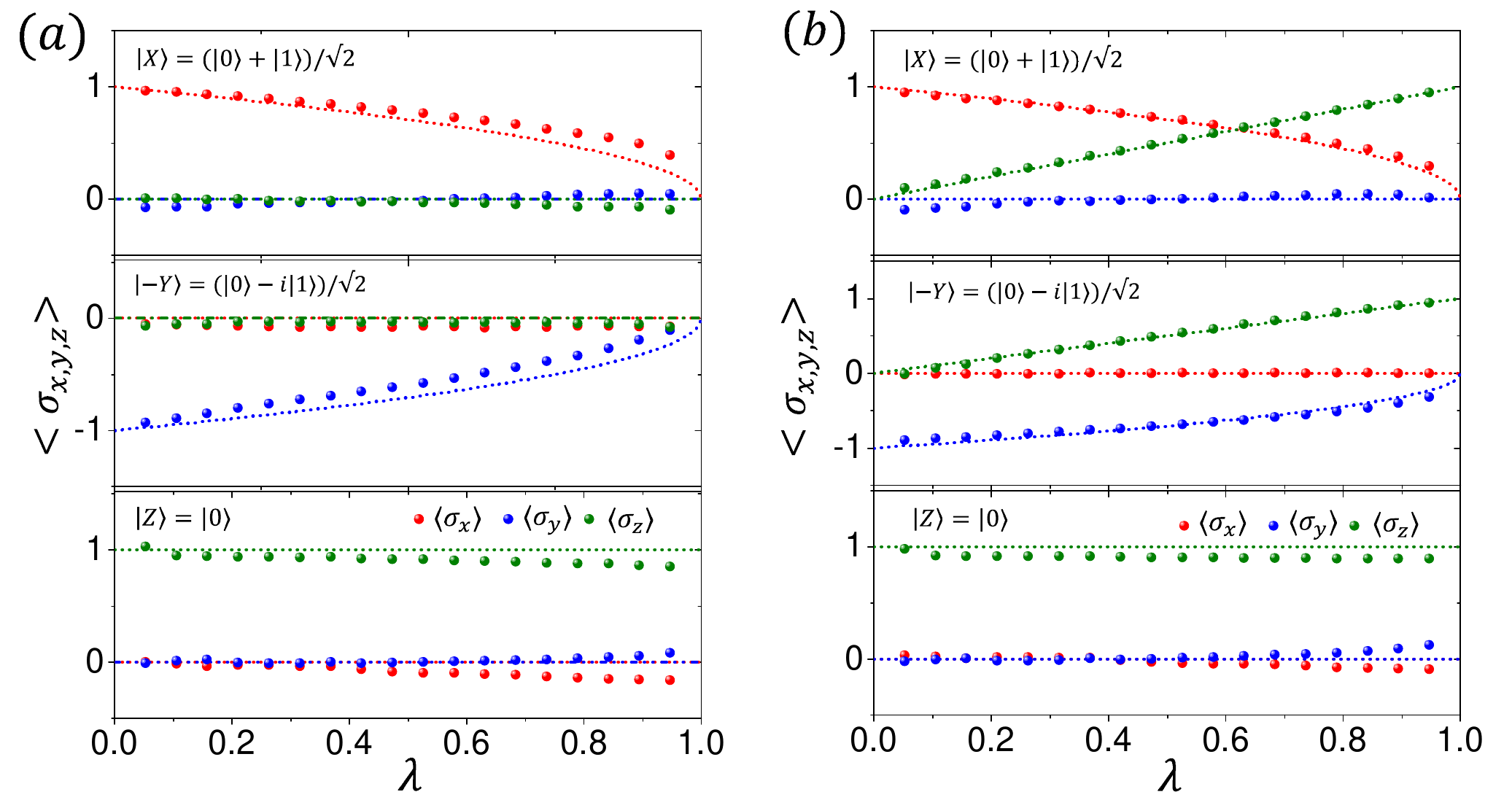}
\end{center}
\setlength{\abovecaptionskip}{-0.00cm}
\caption{\footnotesize{\textbf{Experimental results for the quantum simulation of the PD channel (a), and the AD channel (b).} In each of the two experiments, we initially prepare the working qubit in states $\ket{X}$, $\ket{-Y}$, and $\ket{Z}$, which is easily realised by using a $\pi/2$ pulse around $x$ axis ($y$ axis) starting from the initial state $\ket{Z}=\ket{0}$. Then, we measure the expectation values $\sigma_{x,y,z}$ for the output state of the working qubit. The parameter $\lambda$ is varied from 0 to 1 with 1/20 increments.  }}\label{kraus1}
\end{figure*}

In order to experimentally demonstrate our proposed quantum simulation scheme, we make use of the nuclear spins in a sample of $^{13}$C-labeled chloroform dissolved in deuterated acetone that we manipulate through techniques of NMR\cite{luxin2016,cory2000nmr}. The nuclear spins of $^{13}$C and $^{1}$H are used to encode the two-level working qubit and the ancillary qubit, respectively. The corresponding molecule structure and parameters are illustrated in Fig. \ref{molecule}(a).
Under the weak coupling approximation, the natural Hamiltonian of an $n$-qubit NMR system can be expressed as
 \begin{equation}
\mathcal{H}^n_{int}=\sum_{i=1}^n\omega_i\sigma^i_{z}+\sum_{i<j,i=1}^n\dfrac{\pi J_{ij}}{2}\sigma^i_{z} \sigma^j_{z} \, ,
\end{equation}
where $\omega_i$ is the chemical shift of the $i$th nucleus and $J_{ij}$ is the $J$-coupling constant between the $i$th and the $j$th nuclear spins. 

In experiments, starting from an initial thermal equilibrium state, we first generate a pseudo-pure state (PPS) associated to the state $\ket{0}^{\otimes n}$, as the thermal state is not useful for quantum computation because it is a highly mixed state. For the employed liquid sample, the thermal equilibrium state can be written as 
\begin{equation} 
    \rho_{\text{thermal}}=\frac{\mathcal{I}^{\otimes n}}{2^n}+\sum_{i=1}^{n}\epsilon_i\sigma_z^i \, ,     
\end{equation}
where $n$ is the number of qubits, and $\epsilon_i$ represents the polarization of the $i$-th nucleus at room temperature. The spatial averaging technique was used to initilize our system~\cite{xin2016measurement,knill1998effective,cory1997ensemble}, taking the thermal state to the following PPS 
\begin{equation} 
    \rho_{0}=\frac{1-\epsilon}{2^n}\mathcal{I}+\epsilon\ket{0}\bra{0}^{\otimes n} .   
\end{equation}
A state of this form is convenient as the term related to the identity does not evolve under any unitary propagator and cannot be observed in NMR. Therefore, we can restrict our analysis to the deviation term $\ket{0}\bra{0}^{\otimes n}$ and use it to encode the behaviour of the quantum system. For our experimental analysis, we consider the following initial state for the working qubit: $\ket{X}=(\ket{0}+\ket{1})/\sqrt{2}$, $\ket{-Y}=(\ket{0}-i\ket{1})/\sqrt{2}$, and $\ket{Z}=\ket{0}$. Secondly, for each given input state $\rho_{\text{in}}$, we measure the expectation values <$\sigma_{x,y,z}$>=$\text{Tr}(\rho_{\text{out}}\sigma_{x,y,z})$ on the working qubit at the output of the circuit, after it has undergone all the unitary steps. We do this for a collection of values of the parameter $\lambda$, ranging from $0$ to $1$ and incremental steps of 1/20. The output state of the working qubit $\rho_{\text{out}}$ is directly obtained via single-qubit tomography.

The whole process, from the PPS $\ket{00}\bra{00}$ to the end of the protocol, contains the following steps: a single rotation of the system qubit to prepare its initial state  $\rho_{\text{in}}\otimes \ket{0}\bra{0}$ from the PPS $\ket{00}\bra{00}$, the operations $V$,  $W$, all the controlled operations, and the readout $\pi/2$ pulse. These operations are all packed up together and realised via the GRadient Ascent Pulse Engineering (GRAPE) technique \cite{Khaneja2005,Negrevergne2008}. The GRAPE approach provides a 5ms pulse width and over 99.5\% fidelity for the whole package. Analytically, for any input state of the form $\rho_{\text{in}}=0.5\mathcal{I}+\alpha\sigma_x+\beta\sigma_y+\gamma\sigma_z$, the PD channel should result in a final state $\rho^{\text{th}}_{\text{out}}=0.5\mathcal{I}+\alpha\sqrt{1-\lambda}\sigma_x+\beta\sqrt{1-\lambda}\sigma_y+\gamma\sigma_z$. In Fig. \ref{kraus1}(a), the expectation values of  <$\sigma_{x,y,z}$> are plotted, which agree well with the theoretically expected values. These clearly show that the PD channel reduces all the magnetisation, $M_{x,y}$, in the $xy$ plane, while  keeping the magnetisation, $M_{z}$, in the $z$ direction for any input state $\rho_{\text{in}}$.

{\it{AD channel .--}} We move now to analyse the  case of the AD channel~\cite{fanoc2009}, which is characterised by taking every input to a specific state. The AD channel is described in the Kraus representation via the operators $M_0=[1~0;0~\sqrt{1-\lambda} ]$ and $M_1=[0~\sqrt{\lambda};0~0]$. Alternatively, the AD process can be represented as $M_0\rho_{\text{in}}M_0^\dag+\lambda S_0\rho_{\text{in}}S_0^\dag$,  where $S_0$ is a Kraus operator corresponding to the completely positive and trace preserving process described by the set of Kraus operators $\{S_0, S_1\}$, with $S_0=[0~1;0~0]$ and $S_1=[0~0;1~0]$. For experimental convenience, we choose to implement this second decomposition in terms of $M_0$ and $S_0$. We do this, on the one hand, because the $M_0\rho_{\text{in}}M_0^\dag$ part can be directly obtained from the simulation of the PD channel, and on the other hand, because the simulation of $S_0$ is specially convenient as it does not depend on parameter $\lambda$, and therefore a single experimental run serves to compute the effect of any value of $\lambda$, clearly reducing the experimental requirements. The evolution associated to Kraus operators $S_0$ and $S_1$ can easily be given by the operators $V$, $W$, $U_0$, and $U_1$ taking values
\begin{equation}
\begin{array}{l}
U_0=\sigma_x, U_1=i\sigma_y, V=W=\sqrt{\frac{1}{2}}\left( {\begin{array}{*{20}{c}}
1&1\\
1&-1
\end{array}} \right).
\end{array}
\label{vw2main}
\end{equation}

Thus, the experiment is performed in two steps, corresponding to  the quantum circuit shown in Fig. \ref{circuit1}(a) with two different settings of the operators $V$, $W$, $U_0$, and $U_1$.  The first setting is chosen according to Eq.~(\ref{vw1main}), and only subspace $\ket{0}$ of the ancillary qubit is measured, which is associated to the transformation $M_0\rho_{\text{in}}M_0^\dag$. The second setting is that shown in Eq.~(\ref{vw2main}), and we only measure the subspace of the ancillary qubit corresponding to state $\ket{0}$, which leads to the term $S_0\rho_{\text{in}}S_0^\dag$.  

We use the same sample as that of the previous experiment in order to experimentally simulate the dynamics of the AD channel. We follow the same experimental steps as those in the previous section, performing the experiment twice, for two different setting of the quantum gates in the circuit. As an example, for the case in which the expectation value <$\sigma_y$> of the final state is measured for the initial state $\ket{X}$, we firstly prepare the initial state $\rho^{\text{CH}}_{\text{in}}=\ket{X}\bra{X}\otimes\ket{0}\bra{0}$ from the PPS and drive it following quantum circuit shown in Fig. \ref{circuit1}(a), as described by Eq.~(\ref{vw1main}). Then, the observable $\sigma_{y}\otimes \ket{0}\bra{0}$ is measured to provide the $y$-element associated to the evolution $M_0\rho_{\text{in}}M_0^\dag$. Next, the same preparation and measurement are performed but this time utilizing the setting of unitary operators in Eq.~(\ref{vw2main}), the results corresponding now to the $y$-element associated to $S_0\rho_{\text{in}}S_0^\dag$. Combining these two results, one obtains the desired value <$\sigma_y$> for a qubit undergoing an AD channel. As in the previous experiment, the GRAPE technique is employed to generate the evolution corresponding to the quantum circuits. 
%The whole sequence can be shown to have a 5ms pulse width and a 99.5\% fidelity.

For the input state $\rho_{\text{in}}$, the AD channel can be shown to result in $\rho^{\text{th}}_{\text{out}}=0.5\mathcal{I}+\alpha\sqrt{1-\lambda}\sigma_x+\beta\sqrt{1-\lambda}\sigma_y+(\gamma(1-\lambda)+0.5\lambda)\sigma_z$. In Fig.~\ref{kraus1}(b), we show the experimental measurement of $\langle \sigma_{x,y,z}\rangle$, necessary for the reconstruction of the system qubit, and how these measurement compare to the analytically computed values. The experimental results show a good agreement with the theoretical predictions, which clearly show that the AD channel damps the system towards the ground state $\ket{0}\bra{0}$, reducing the magnetisation in the $xy$ plane, while increasing it in the $z$ direction. This could be of interest in the initialisation of a system that is in an arbitrary state.

\begin{figure}[htb]
\begin{center}
\includegraphics[width= 0.9\columnwidth]{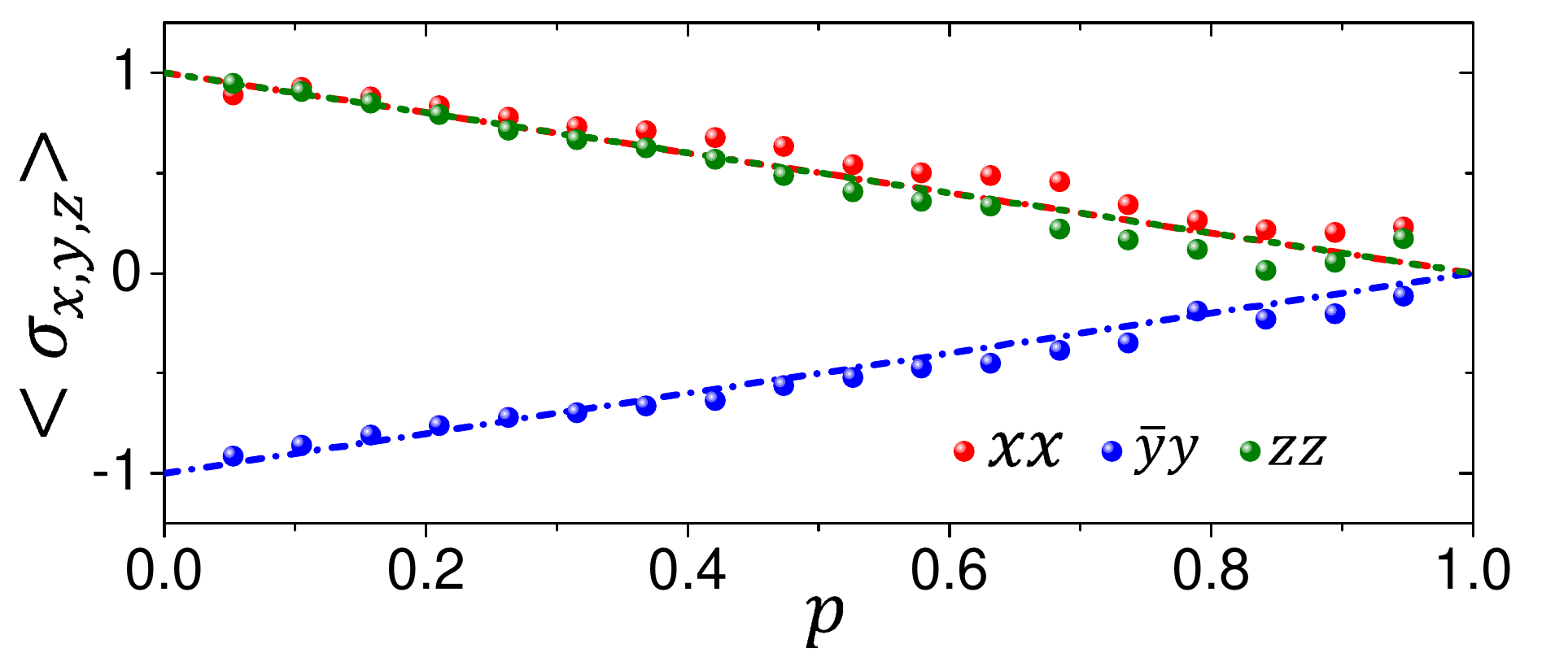}
\end{center}
\setlength{\abovecaptionskip}{-0.00cm}
\caption{\footnotesize{\textbf{Experimental results for the quantum simulation the DEP channel. } We perform three experiments, preparing the system qubit in state $+x$, $-y$, and $+z$, which corresponds to a direction in the Bloch sphere. We then measure the expectation value of a Pauli matrix in the same axis for each initial state.  For instance, the observable <$\sigma_y$>=Tr$(\rho_{\text{out}}\sigma_y)$ will be measured if the initial state is  $-y$ direction. }}\label{kraus2}
\end{figure}

{\it{DEP channel .--}} To complete our study of decoherence channels, we consider the DEP channel $\large \varepsilon^{\text{DEP}}$~\cite{Bennett1999}. For this case, the Kraus representation is given by $E_0=\sqrt{1-\frac{3p}{4}}\mathcal{I}$, $E_1=\sqrt{\frac{p}{4}}\sigma_x$, $E_2=\sqrt{\frac{p}{4}}\sigma_y$, and $E_3=\sqrt{\frac{p}{4}}\sigma_z$. It can be trivially shown that the effect of the DEP channel on an initial state $\rho_{\text{in}}$ is to evolve towards the maximally mixed state $\large \varepsilon^{\text{DEP}}(\rho_{\text{in}})=p\mathcal{I}/2+(1-p)\rho_{\text{in}}$, with some probability $p$. 

In Fig.~\ref{circuit1}(b), the quantum circuit to realise the quantum simulation of the DEP channel is depicted, following our proposed protocol.  In this case, we set $U_0=\mathcal{I}$, $U_1=\sigma_x$, $U_2=\sigma_y$, and $U_3=\sigma_z$, while operator $V$ is given by
\begin{equation}
\left( {\begin{array}{*{20}{c}}
\sqrt{1-\frac{3p}{4}}&-\sqrt{\frac{\frac{p}{4}(1-\frac{3p}{4})}{1-\frac{p}{4}}}&-\sqrt{\frac{\frac{p}{4}(1-\frac{3p}{4})}{(1-\frac{p}{4})(1-\frac{p}{2})}}&-\sqrt{\frac{p}{4-2p}}\\
\sqrt{\frac{p}{4}}&\sqrt{1-\frac{p}{4}}&0&0\\
\sqrt{\frac{p}{4}}&-\frac{p}{4\sqrt{1-\frac{p}{4}}}&\sqrt{\frac{1-\frac{p}{2}}{1-\frac{p}{4}}}&0\\
\sqrt{\frac{p}{4}}&-\frac{p}{4\sqrt{1-\frac{p}{4}}}&-\frac{p}{4\sqrt{(1-\frac{p}{4})(1-\frac{p}{2})}}&\sqrt{\frac{4-3p}{4-2p}}
\end{array}} \right) . \nonumber
\end{equation}
On the other hand, operation $W$ is fixed to a $4\times4$ identity matrix. The quantum circuit for the DEP channel is then implemented by combining two ancillary qubits initially prepared in the state $\ket{00}\bra{00}$ and a system qubit in input state $\rho_{\text{in}}$.  At the end of the protocol, the two ancillary qubits are traced out to acquire the output state of the DEP channel $\rho_{\text{out}}=\large \varepsilon^{\text{DEP}}(\rho_{\text{in}})$. 

Experimentally, we need a three-qubit quantum-information processor, which is implemented via diethyl fluoromalonate dissolved in d6 acetone in NMR, where the nuclear spins of $^{13}$C, $^{1}$H, and $^{19}$F in the diethyl fluoromalonate molecule act as the system qubit and the two ancillary qubits, respectively. Figure \ref{molecule}(b) shows the corresponding structure and parameters. The spatial averaging technique is again used to prepare the PPS $\ket{000}\bra{000}$ \cite{xine2015}. The unitary operators are implemented via the GRAPE technique that provides a 10ms pulse width. For the DEP channel, we only carry out $xx$-, $\bar{y}y$-, and $zz$-experiments, which are enough to demonstrate the properties of the DEP channel. For instance, for the case of the $\bar{y}y$-experiment, we prepare the system in the state $\ket{-Y}\bra{-Y}\otimes\ket{00}\bra{00}$ by applying a $\pi/2$ pulse around the $x$ axis to the system qubit $^{13}$C. Then, we measure the observable $\sigma_y\otimes \mathcal{I}\otimes \mathcal{I}$ which provides us with the expectation value <$\sigma_y$>=Tr$(\rho_{\text{out}}\sigma_y)$. $\rho_{\text{out}}$ is the output state of the system qubit $^{13}$C after tracing out the ancillary qubits $^{1}$H and $^{19}$F at the end of quantum circuit. Two other experiments are performed in a similar fashion similarly corresponding to $\bar{y}y$-experiments. Fig.~(\ref{kraus2}) illustrates the corresponding results of three experiments $xx$, $\bar{y}y$, and $zz$ for different values of $p$, which presents a good agreement between the theoretical predictions and the experiments.

\begin{figure*}[htb]
\begin{center}
\includegraphics[width= 1.7\columnwidth]{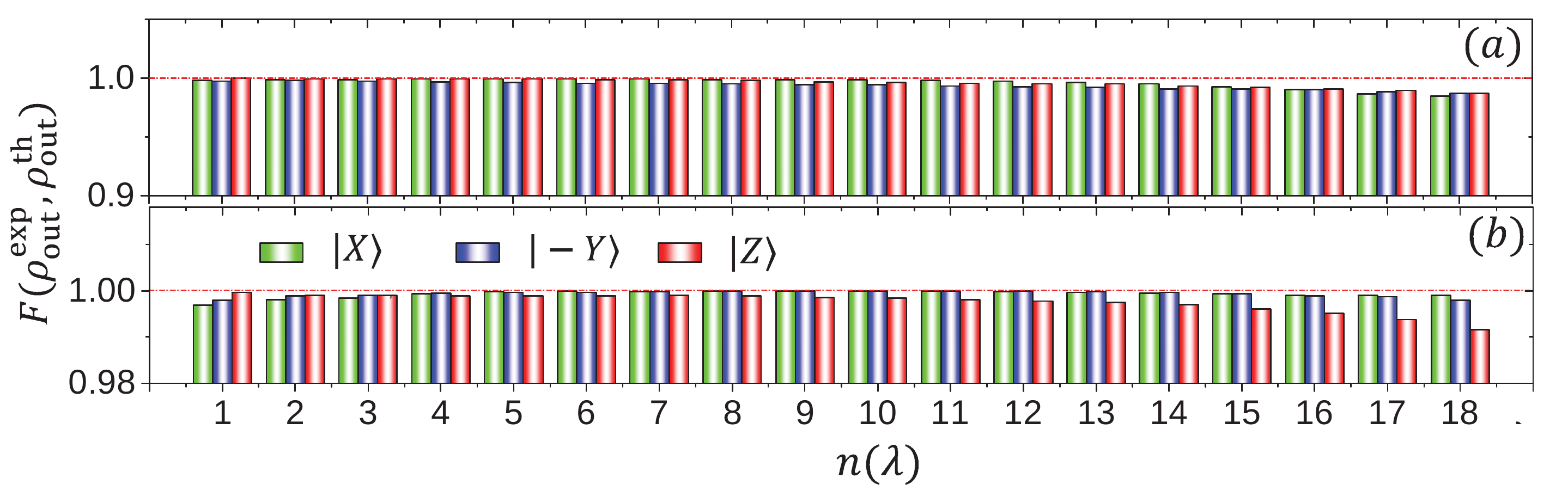}
\end{center}
\setlength{\abovecaptionskip}{-0.00cm}
\caption{\footnotesize{\textbf{Fidelity $F(\rho^{\text{exp}}_{\text{out}},\rho^{\text{th}}_{\text{out}})$ of the state after the quantum simulation of PD and AD channel with respect to the analytically expected state.}  For a number of input states $\ket{X}$, $\ket{-Y}$, and $\ket{Z}$, the output density matrices $\rho^{\text{exp}}_{\text{out}}$ are measured and their fidelity $F(\rho^{\text{exp}}_{\text{out}},\rho^{\text{th}}_{\text{out}})$ with respect to the ideal values is computed. The subfigure ($a$) and  ($b$) present the corresponding fidelities for the PD and AD channels, respectively. $n(\lambda)$ represents each of the steps of parameter $\lambda$ as it increases from 0 to 1 in 18 steps.}}\label{F_detail}
\end{figure*}

\section{Discussion}
In order to evaluate the accuracy of our simulations, we have computed the fidelity $F(\rho^{\text{exp}}_{\text{out}},\rho^{\text{th}}_{\text{out}})$ between the reconstructed single-qubit density matrix $\rho^{\text{exp}}_{\text{out}}$ and the ideal state $\rho^{\text{th}}_{\text{out}}$ for each of the PD and AD channels using the following  procedure. These results are illustrated in Fig.~\ref{F_detail}. The adopted fidelity definition through the whole work is $F=\text{Tr}(\rho_a\rho_b)/\sqrt{\text{Tr}(\rho_a^2)\text{Tr}(\rho_b^2)}$. In our experiments, the average fidelities between the reconstructed single-qubit state $\rho^{\text{exp}}_{\text{out}}$ and the ideal output state $\rho^{\text{th}}_{\text{out}}$ are around 99.52\% and 99.87\% for the PD and AD channels, respectively.

For an $n$-qubit quantum system $\rho_{\text{SA}}$, with one system qubit and $(n-1)$ ancillary qubits,  the operation of tracing out the ancillary qubits, in order to obtain the state of our system qubit $\rho_{\text{S}}=\text{Tr}_\text{A}(\rho_{\text{SA}})$, can be realised by measuring the following operators of the output state $\rho_{\text{SA}}$,
\begin{equation}
\mathcal{M}_{x,y,z}=\sigma_{x,y,z}\otimes \mathcal{I}^{\otimes n-1}.
\end{equation}
In an NMR platform, $2^{n-1}$ peaks will be observed, with the $m$-th peak providing the expectation values of operators
\begin{equation}
\begin{array}{l}
\mathcal{M}_x^{m,n}=\sigma_{x} \otimes \ket{b(m-1,n-1)}\bra{b(m-1,n-1)}\ \ {\rm and}\\
\mathcal{M}_y^{m,n}=\sigma_{y} \otimes \ket{b(m-1,n-1)}\bra{b(m-1,n-1)},
\end{array}
\end{equation}
where $b(m-1,n-1)$ is the binary representation of number $m-1$ in $n-1$ bits. 

Summing the following results over $m$ from 1 to $2^{n-1}$ leads to $\mathcal{M}_{x,y}=\sum_{m=1}^{2^{n-1}} \mathcal{M}_{x,y}^{m,n}$. To measure the observable $\mathcal{M}_{z}$, we apply an additional readout pulse ($\pi/2$ pulse around $y$ axes) on the system qubit at the end, which transfers the magnetization in the $z$ direction to the $x$ direction. In this manner, the expectation value of $\mathcal{M}_{x}$ corresponds the value of the desired  observable $\mathcal{M}_{z}$. Moreover, single-qubit tomography of the system qubit can easily be realised using the following rule,
\begin{equation}
\rho_{\text{S}}=\frac{1}{2}\mathcal{I}+\frac{\left \langle \mathcal{M}_{x} \right \rangle}{2^n}\sigma_x+\frac{\left \langle \mathcal{M}_{y} \right \rangle}{2^n}\sigma_y+\frac{\left \langle \mathcal{M}_{z} \right \rangle}{2^n}\sigma_z,
\end{equation}
where the coefficient $2^n$ is a normalisation constant, and $\left \langle \mathcal{M}_{x,y,z} \right \rangle$ is the expectation value of observable $ \mathcal{M}_{x,y,z}$, $\text{Tr}(\rho_{\text{SA}}\mathcal{M}_{x,y,z})$.
\\ 

\begin{figure}[htb]
\begin{center}
\includegraphics[width= 0.92\columnwidth]{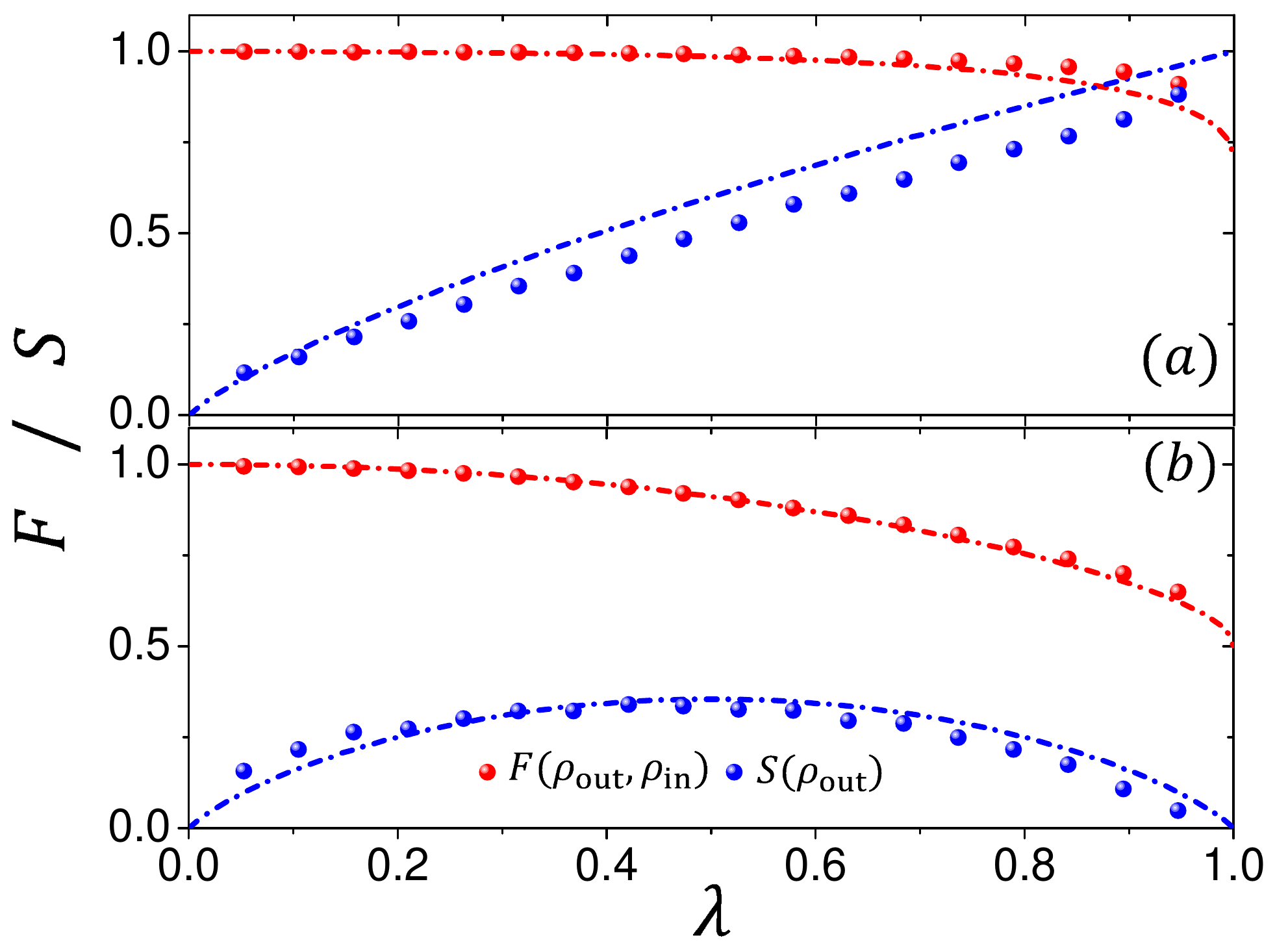}
\end{center}
\setlength{\abovecaptionskip}{-0.00cm}
\caption{\footnotesize{\textbf{Fidelity $F(\rho_{\text{out}},\rho_{\text{in}})$ and entropy $S(\rho_{\text{out}})$ for a qubit evolving under a PD channel (a), and an AD channel (b).}  $F(\rho_{\text{out}},\rho_{\text{in}})$ is decreasing because the output $\rho_{\text{out}}$ slowly deviates from the input state $\rho_{\text{in}}$. The behaviour of $S(\rho_{\text{out}})$ is directly guided by the entanglement strength between the system qubit and the environment, or the purity of the system qubit after tracing out the environment. }}\label{FS}
\end{figure}

We further complete our analysis by the study of the behaviour of some additional properties under these quantum channels. More precisely,  we look at the fidelity $F(\rho_{\text{out}},\rho_{\text{in}})$ and the von Neumann entropy $S(\rho_{\text{out}})=-\text{Tr}(\rho_{\text{out}}\text{log}_2\rho_{\text{out}})$, for an input state $\ket{X}$ as it undergoes the PD and AD channels. These results are illustrated in Fig.~(\ref{FS}). $F(\rho_{\text{out}},\rho_{\text{in}})$ reflects the strength of the quantum channel acting on a qubit, which decreases for higher strength $\lambda$. $S(\rho_{\text{out}})$ quantifies the strength of the entanglement between the system qubit and the ancillary system, such that $S$ will increase together with the strength $\lambda$ of the PD channel. On the opposite, for the AD channel the entropy clearly shows a maximum for $\lambda=0.5$, while it vanishes for minimal ($\lambda=0$) and maximal ($\lambda=1$) values of $\lambda$. This happens because under the PD channel the initial state $\ket{X}$ will tend towards a maximally mixed state while, for an AD channel, it will gradually tend towards the ground state $\ket{0}$ through intermediate mixed states, respectively.
\\

\section*{Conclusion}

We have proposed a method for the quantum simulation of open quantum dynamics and experimentally implemented the proposed simulation procedure, realizing proof-of-principle experiments in an NMR setup. Our experiment is a small-scale demonstration of the working principles of the proposed techniques, which can be considered as building blocks for more involved protocols.  The experimental results show a high degree of correspondence with the theoretical predictions, showing the capacity of our method to simulate paradigmatic decoherence channels. A natural extension of this work is the development of  methods to construct algorithms to simulate the dynamics of open quantum systems in higher dimensions. For example, using the Weyl operator basis, any three-dimensional channel in the Kraus representation can  be decomposed into a linear combination of Weyl operators: $M= \sum_{n,m=0}^{2} b_{nm} U_{nm}$ \cite{welyl}, where  $U_{nm} $ is the  Weyl operator and $ b_{nm} $ are coefficients. Namely, we can perform any Kraus operator in the form of a linear combination of Weyl operators with the proposed method.

\begin{acknowledgments}
T. X. , S. W. and G. L. are grateful to the following funding sources: National Natural Science Foundation of China under Grants No. 11175094 and No. 91221205; National Basic Research Program of China under Grant No. 2015CB921002. J. S. P. and E. S. acknowledge financial support from grants: Spanish MINECO/FEDER FIS2015-69983-P and Basque Government IT986-16.
\end{acknowledgments}

 \section*{Appendix A: Scaling of the Protocol}

 The size of the ancillary system in our protocol is given by the greatest of these two: the number of Kraus operators $d_1$, and the number of unitary operators $d_2$ onto which the Kraus operators are decomposed. For an $n$-qubit system, with a Hilbert space dimension $d_S=2^n$, any operator can be decomposed in the Weyl basis~\cite{welyl} as the complex superposition of maximally $d_S^2$ unitary operators, which are also traceless, and trace-wise orthogonal. Therefore, all the simulated Kraus operators, which act on a system of $n$ qubits, can be decomposed into a basis of not more than $N=2^{2n}$ unitary operators.  As a consequence, the total number of ancillary qubits is upper bounded by $\log_2(N)=2n$. This is similar to other simulation approaches, like for example the Stinespring dilatation method, which also takes a maximum of $2n$ ancillary qubits.  
 
In order to count the number of required gates, we split our protocol in two parts. On the one hand, we have the initial and final operations $V$ and $W$, which act on the ancillary system and are, in general, arbitrary matrices. It is known that an arbitrary unitary operation acting on an $M$-qubit system can always be implemented with a circuit containing a total of $O(M^3 2^{2M})$ single qubit and CNOT gates~\cite{N, complex1, complex2}. Therefore, in the most unfavourable case, where a total of $2n$ ancillary qubits are required, our method would employ up to $O(8n^32^{4n})$ single-qubit and two-qubit gates to implement the $V$ and $W$ operations. 

On the other hand, we have the controlled unitary operations acting on the target system of $n$ qubits. This operations are not arbitrary, but they correspond to a specific basis of unitary operators. One can, for example, choose a basis consisting of the tensor product of Pauli operators. 
\begin{figure}[htb]
\begin{center}
\includegraphics[width= 0.92\columnwidth]{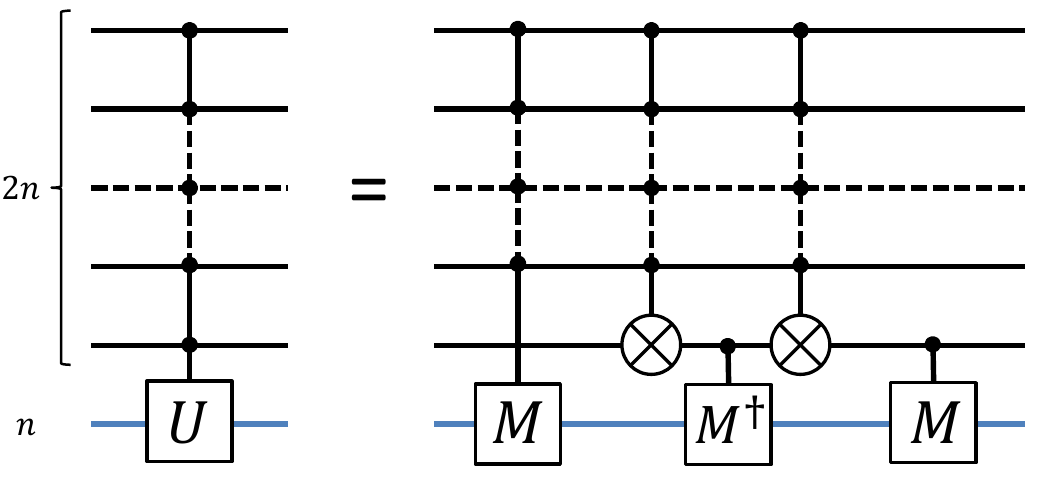}
\end{center}
\setlength{\abovecaptionskip}{-0.00cm}
\caption{\footnotesize{\textbf{The decomposition of a controlled operation $C_{2n}(U)$.}  $C_{2n}(U)$ can be decomposed into a combination of controlled operations $C_{2n-1}(M)$, $C_{1}(M^\dagger)$, $C_{1}(M)$ and two Toffolli gates over $2n$ qubits. Here, $M^2=U$. }}\label{gate}
\end{figure}
In this case, it can be shown that the gate complexity for each of the controlled operations goes like $O(n^2)$~\cite{complex1}. Let $C_{m}(U)$ denote a controlled gate where the number of control qubits is $m$ and $U$ acts on a target system of $n$ qubits. We will use $T_m$ to denote the gate cost of decomposing $C_{m}(U)$. The circuit in Fig.~\ref{gate} shows a suitable decomposition of $C_{2n}(U)$. Moreover, the Toffolli gate over $2n$ qubits can be decomposed into $O(n)$ single-qubit and CNOT gates. On the other hand, $M^\dagger$ and $M$, which fulfil $M^2=U$, can also be decomposed into $n$ single-qubit gates, as $U$ is a tensor product of Pauli matrices.  The cost of decomposing the gates $C_{1}(M)$ and $C_{1}(M^\dagger)$ is therefore $O(n)$. From such a decomposition, the following recurrence relation can be inferred,
\begin{equation}
T_m=T_{m-1}+O(n). 
\end{equation}
Hence, the total gate complexity to implement each controlled unitary operator $C_{2n}(U)$ is proportional to $O(n^2)$.

According to the discussion above, a total of $2^{2n}$ unitary operations form a complete basis of the $n$-qubit system, and therefore the implementation of these basis operators controlled with respect to the ancillary system takes a total of $O(n^2 2^{2n})$ single qubit and CNOT gates. In total our algorithm in the most general case can be associated to a gate complexity of $O(8n^3 2^{4n}+n^2 2^{2n})$. If we consider other simulation methods, like the Stinespring dilatation, were the system is enlarged to accommodate a $2n$-qubit environment in the most general case, we find that to perform an arbitrary unitary operation in the enlarged Hilbert space, we need up to  $O(27n^3 2^{6n})$ single qubit and CNOT gates, which is an exponential factor more gates than we need. This is because, while our method keeps the size of the operations either to the size of the system or to that of the ancillary space, the Stinespring dilatation method needs to perform operations on the complete system-plus-ancilla space. However, for specific cases where the number of Kraus operators is small and their decomposition requires a large number of unitary operators, it  can be the case that the Stinespring dilatation method is more convenient. Ultimately, the comparison should be done case by case.
 
 \section*{Appendix B: Measurements in NMR}
 
 While NMR spectroscopy is a so-called ensemble weak measurement, which does not collapse the total wave function, expectation values of arbitrary global spin observables can be measured, and with these one can reproduce the outcome of projective measurements, which can be distinguished by the spectra of the NMR ensamble and individually operated on with selective pulses in NMR. In this manner, one can imitate the outcomes of projective measurements and their associated   probabilities~\cite{lee,auccaise,laflamme}. Besides, in NMR, the measurement of the expectation value of an observable corresponds to the spectroscopy of macroscopic ensembles of quantum spins, which results in a usually significantly precise and stable measurement. Indeed, the precision of the measured data is such that the error bars are typically smaller than the plotted dots, as it is the case for the experimental data presented throughout this paper.  
 
   \begin{table}[tbp]
\centering
\caption{\footnotesize{The standard deviations between simulated results and theoretical predictions. The subscript in $\epsilon$ indicates the input state.}} \label{com}
\begin{tabular}{cccc}
\hline
\hline
Deviations & $\epsilon_x$ & $\epsilon_{\bar{y}}$ & $\epsilon_z$ \\
\hline
PD channel&0.0136 & 0.0158 &0.0107\\
AD channel& 0.0091 &0.0098&0.0103\\
DEP  channel&0.0994&0.0203&0.0434\\
\hline
\hline
\end{tabular}
\end{table}

Finally, the minor deviations of the measured data can be associated to imperfections of the PPS initialization, imprecisions of the GRAPE pulses, and dephasing effects caused by decoherence, which are the leading sources of error in our setup. We have numerically simulated the GRAPE pulses including a contrasted decoherence model for our qubits, in order to estimate an error bar for each simulated channel. We compute the standard deviation of our simulated data as $\epsilon=\sqrt{\sum_{i=1}^M(x^i_{\text{sim}}-x^i_{\text{th}})^2/(M-1)}$, with $M$ the number of sampling points. In Tab.~\ref{com}  we give the results for different input states in the PD, AD, and DEP channels.

\end{document}